\documentclass[10pt,conference]{IEEEtran}

\usepackage{amsmath}
\usepackage{amssymb}
\usepackage{stmaryrd}
\usepackage[T1]{fontenc}
\usepackage{url}
\newcommand{\abs}[1]{\left| #1 \right|}
\newcommand{\aabs}[1]{\left\| #1 \right\|}
\newcommand{\okra}[1]{\left( #1 \right)}
\newcommand{\kwad}[1]{\left[ #1 \right]}
\newcommand{\klam}[1]{\left\{ #1 \right\}}

\newcommand{\floor}[1]{\left\lfloor #1 \right\rfloor}
\newcommand{\dzi}[1]{\left\langle #1 \right\rangle}
\newcommand{\sred}[1]{\textbf{E}\, #1}
\newcommand{\voc}[1]{\textbf{V}[#1]}

\DeclareMathOperator{\card}{card}

\newtheorem{definition}{Definition}

\newtheorem{theorem}{Theorem}
\newtheorem{lemma}{Lemma}

\begin{document}

\title{On vocabulary size of grammar-based codes}

\author{
\authorblockN{{\L}ukasz D\k{e}bowski}
\authorblockA{Institute of Computer Science\\
  Polish Academy of Sciences\\
  01-237 Warszawa, Poland\\
  Email: ldebowsk@ipipan.waw.pl}
}

\maketitle

\begin{abstract}
  We discuss inequalities holding between the vocabulary size, i.e.,
  the number of distinct nonterminal symbols in a~grammar-based
  compression for a~string, and the excess length of the respective
  universal code, i.e., the code-based analog of algorithmic mutual
  information.  The aim is to strengthen inequalities which were
  discussed in a~weaker form in linguistics but shed some light on
  redundancy of efficiently computable codes.  The main contribution
  of the paper is a~construction of universal grammar-based codes for
  which the excess lengths can be bounded easily.
\end{abstract}

\IEEEpeerreviewmaketitle

\section{Introduction}
\label{secIntro}

In recent years some interest in the theory of universal coding has
focused on detecting hierarchical structure in compressed data.  An
important tool for this task are universal grammar-based codes
\cite{KiefferYang00} which compress strings by transforming them first
into special context-free grammars \cite{CharikarOthers05} and then
encoding the grammars into less redundant strings.  This article
presents several bounds for the vocabulary size, i.e., the number of
distinct nonterminal symbols in a~grammar-based compression for
a~string.  Indirectly, the bounds concern also the code redundancy,
which can be elucidated as follows.

Let $X_{m:n}:=(X_k)_{m\le k\le n}$ be the blocks of finitely-valued
variables $X_i:\Omega\rightarrow\mathbb{X}=\klam{0,1,...,D-1}$ drawn
from stationary process $(X_k)_{k\in\mathbb{Z}}$ on
$(\Omega,\mathcal{J},P)$. Assuming expectation operator $\sred{}$,
define $n$-symbol block entropy $H(n):=H(X_{1:n})=-\sred{\log
  P(X_{1:n})}$ and excess entropy $E(n):=
I(X_{1:n};X_{n+1:2n})=2H(n)-H(2n)$, being mutual information between
adjacent blocks \cite{CrutchfieldFeldman01}.

On the other hand, let $C:\mathbb{X}^+\rightarrow \mathbb{X}^+$ be
a~uniquely decodable code. For code length $\abs{C(\cdot)}$ being an
analog of algorithmic complexity \cite{CharikarOthers05}, define
$$I^C(u:v):=\abs{C(u)}+\abs{C(v)}-\abs{C(uv)}$$ as the analog of
algorithmic mutual information \cite{GrunwaldVitanyi03}.  We
will denote the expected normalized code length and its excess as
\begin{align*}
  H^C(n)&:=\sred{\abs{C(X_{1:n})}} \log D,
  \\
  E^C(n)&:=\sred{I^C(X_{1:n}:X_{n+1:2n})} \log D .
\end{align*}

For a~uniquely decodable code, noiseless coding inequality $H^C(n)\ge
H(n)$ is satisfied and the code is called universal if compression
rate $\lim_{n} H^C(n)/n$ equals entropy rate $h:=\lim_{n} H(n)/n$ for
any stationary distribution $P((X_k)_{k\in\mathbb{Z}}\in\cdot)$.  In
fact, the search for codes having the lowest redundancy on finite
strings can be restated as the task of finding universal codes with
the smallest excess code length $I^C(\cdot:\cdot)$ since
  \begin{align}
   \label{DiffECE}
    \limsup_{n\rightarrow\infty} & \kwad{E^C(n) - E(n)}\ge 0,
    \\
    \limsup_{n\rightarrow\infty} & \kwad{E^C(n) - E^{C'}(n)}\ge 0
    \text{ if $H^C(\cdot)\ge H^{C'}(\cdot)$},
\end{align}
for any universal codes $C$ and ${C'}$, cf.\ \cite{Debowski06,Debowski06c}.  

The specific aim of the present note is to justify links between the
vocabulary size and excess code length $I^C(\cdot:\cdot)$ for
certain universal grammar-based codes.  A~weaker form of this
connection was mentioned in the context of following linguistic
investigations, cf.\ \cite{Debowski06,Debowski06b}:
\begin{LaTeXenumerate}
\item Majority of words in a natural language text can be identified
  as frequently repeated strings of letters. Grammar-based codes can
  be used to detect these repeats. Distinct words of the text happen
  to get represented as distinct nonterminal symbols in an
  approximately smallest context-free grammar for the text
  \cite{Wolff80,DeMarcken96}.  The number of different
  ``significantly'' often repeated substrings in a~typical text can be
  100 times greater than in a~comparable realization of a~memoryless
  source \cite{Debowski06b}.
\item There is a~hypothesis that excess entropy of a~random natural
  language text (imagined as a~stationary stochastic process with
  $X_i$ being consecutive letters of the text) obeys $E(n)\asymp
  \sqrt{n}$ rather than $E(n)=0$ as for a~memoryless source
  \cite{Hilberg90} (cf.\ \cite{Debowski06c} for a~connection of such
  an effect with nonergodicity). We asked whether the power-law growth
  of $E(n)$ can be linked with the known empirical power-law growth of
  the number of distinct words in a~text against the text length
  \cite{Herdan64}.
\end{LaTeXenumerate}
In view of observation (i), our question in (ii) could be restated as:
Are excess entropy $E(n)$ and the expected vocabulary size of some
minimal code for string $X_{1:2n}$ approximately equal for every
stationary process? Trying to answer the question, we derived
inequality (\ref{DiffECE}) in \cite{Debowski06} and sought for further
links between the excess code length and the vocabulary size.  The
result of \cite{Debowski06} concerning the latter is encouraging but
too weak.  It relates the vocabulary size of the smallest grammar in
the sense of \cite{CharikarOthers05} to the Yang-Kieffer excess
grammar length rather than to the excess length of an actual universal
code. 

In this article, we will strengthen the connection.  We will prove
that excess code length $I^C(u:v)$ for some grammar-based code $C$ is
dominated by the product of the length of the longest repeated
substring in string $w:=uv$ and the vocabulary size of the code for
$w$. To get this inequality, it suffices that $C$ be the shortest
code in an algebraically closed subclass of codes using a~special
grammar-to-string encoder. There exist universal codes satisfying
this requirement.

Besides the mentioned dominance, we will justify an inequality in the
opposite direction and, additionally, show that the vocabulary size of
an irreducible grammar for string $w$ cannot be less than the square
root of the grammar length, cf.\ \cite{Debowski06b,KiefferYang00}.
This pair of inequalities might be used to lower-bound the redundancy
of codes based on irreducible grammars.

The exposition is following. Section \ref{secMorph} reviews
grammar-based coding.  We construct local grammar-to-string encoders
(\ref{ssecEncoders}) and define minimal codes (\ref{ssecLengths}) with
respect to some classes of grammars (\ref{ssecGrammars}). Subsection
\ref{ssecUniversal} justifies universality of certain minimal codes
which use local encoders.  Section \ref{secVocabulary} presents the
upper (\ref{ssecUpperExcess}) and the lower (\ref{ssecLowerExcess})
bounds for the excess lengths of a~minimal code expressed in terms of
its vocabulary size.  Section \ref{secConclude} resumes the article.

\section{Grammar-based coding revisited}
\label{secMorph}

Grammar-based compression is founded on the following concept.  An
\emph{admissible} grammar is a~context free-grammar which generates
singleton language $\klam{w}$, $w\in\mathbb{X}^+$, and whose
production rules do not have empty right-hand sides
\cite{KiefferYang00}. In such a~grammar, there is one rule per
nonterminal symbol and the nonterminals can be ordered so that the
symbols are rewritten onto strings of strictly succeeding symbols
\cite{KiefferYang00}.

Hence, an admissible grammar is given by its set of production
rules 
$
\klam{
A_1\rightarrow\alpha_1,
A_2\rightarrow\alpha_2,
...,
A_n\rightarrow\alpha_n 
}$,
where $A_1$ is the start symbol, other $A_i$ are 
secondary nonterminals, and the right-hand sides of rules satisfy
$\alpha_i\in (\klam{A_{i+1},A_{i+2},...,A_n}\cup\mathbb{X})^+$.  Since
the grammar can be restored also from sequence
\begin{align}
\label{GrammarNotation}
  G=(\alpha_1,\alpha_2,...,\alpha_n),
\end{align}
we will call $G$ simply the grammar. Its \emph{vocabulary size}, i.e.,
the number of used nonterminal symbols, will be written
$$\voc{G}:=\card \klam{A_{1},A_{2},...,A_n} =n.$$

Let $\mathbb{X}^*=\mathbb{X}^+\cup\klam{\lambda}$, where $\lambda$ is
the empty word.  For any string $\alpha\in
(\klam{A_{2},A_{3},...,A_n}\cup\mathbb{X})^*$, we denote its
\emph{expansion} with respect to $G=(\alpha_1,\alpha_2,...,\alpha_n)$
as $\dzi{\alpha}_G$ \cite{CharikarOthers05}, i.e.,
$\klam{\dzi{\alpha}_G}$ is the language generated by grammar
$(\alpha,\alpha_2,\alpha_3,...,\alpha_n)$.  The set of admissible
grammars will be denoted as $\mathcal{G}$ and $\mathcal{G}(w)$ will be
the subset of admissible grammars which generate language $\klam{w}$,
$w\in\mathbb{X}^+$.  Function
$\Gamma:\mathbb{X}^+\rightarrow\mathcal{G}$ such that
$\Gamma(w)\in\mathcal{G}(w)$ for all $w\in\mathbb{X}^+$ is called
a~\emph{grammar transform} \cite{KiefferYang00}.

If string $w$ contains many repeated substrings then some grammar in
$\mathcal{G}(w)$ can ``factor out'' the repetitions and may be used to
represent $w$ concisely. It is not straightforward, however, how to
quantify the size of a~grammar. In \cite{KiefferYang00} the length of
grammar $G=(\alpha_1,\alpha_2,...,\alpha_{\voc{G}})$ was defined as
\begin{align}
\label{YKlength}
\abs{G}:=\textstyle\sum_i \abs{\alpha_i},
\end{align}
where $\abs{\alpha}$ is the length of $\alpha\in
(\klam{A_{1},A_{2},...,A_n}\cup\mathbb{X})^*$. Function
(\ref{YKlength}) will be called Yang-Kieffer length.

For a grammar transform, ratio $\abs{\Gamma(w)}/\abs{w}$ can be quite
a~biased measure of string compressibility. Precisely, transform
$\Gamma$ is called \emph{asymptotically compact} if
\begin{align}
\label{AC}
\lim_{n\rightarrow\infty} \max_{w\in\mathbb{X}^n} \abs{\Gamma(w)}/n=0
\end{align}
and for each grammar in $\Gamma(\mathbb{X}^+)$ each nonterminal has
a~different expansion.  There is plenty of such transforms
\cite{KiefferYang00,CharikarOthers05}.

Since the compression given by (\ref{AC}) is apparent, consider
\emph{grammar-based codes}, i.e., uniquely decodable codes
$C=B(\Gamma(\cdot)):\mathbb{X}^+\rightarrow \mathbb{X}^+$, where
$\Gamma:\mathbb{X}^+\rightarrow\mathcal{G}$ is a~grammar transform and
$B:\mathcal{G}\rightarrow\mathbb{X}^+$ is called a~\emph{grammar
  encoder} \cite{KiefferYang00}. We have $\lim_n
\max_{w\in\mathbb{X}^n} \abs{C(w)}/n\ge 1$ necessarily.  Nevertheless,
there exists a~grammar encoder
$B_\text{YK}:\mathcal{G}\rightarrow\mathbb{X}^+$ \cite{KiefferYang00}
such that
\begin{LaTeXenumerate}
\item set $B_\text{YK}(\mathcal{G})$ is prefix-free,
\item $\abs{B_\text{YK}(G)}\le \abs{G}(A+\log_D \abs{G})$ for some
  $A>0$,
\item $C=B_\text{YK}(\Gamma(\cdot))$ is a~universal code for any
  asymptotically compact transform $\Gamma$.
\end{LaTeXenumerate}

\subsection{Local grammar encoders}
\label{ssecEncoders}

It is hard to analyze the excess lengths of grammar-based codes which
use $B_\text{YK}$ given by \cite{KiefferYang00} as their
grammar-to-string encoder. We will define a~more convenient encoder.
It will represent a~grammar as a~string resembling list
(\ref{GrammarNotation}) but, simultaneously, it will constitute nearly
a~homomorphism between some operations on grammars and strings.

\begin{definition} 
$\oplus:\mathcal{G}\times\mathcal{G}\rightarrow\mathcal{G}$
is called \emph{grammar joining} if 
$$
 G_1\in\mathcal{G}(w_1) \land
 G_2\in\mathcal{G}(w_1)\implies
G_1\oplus G_2\in\mathcal{G}(w_1w_2).
$$  
\end{definition}
\null 
It would be convenient to use such grammar joining $\oplus$ and
encoder $B:\mathcal{G}\rightarrow\mathbb{X}^+$ that the edit
distance between $B(G_1\oplus G_2)$ and $B(G_1)B(G_2)$ be small.
Without making the idea too precise, such joining and encoder will be
called \emph{adapted}.

The following example of mutually adapted joining $\oplus$ and
encoders will be used in the next sections.  For any function
$f:\mathbb{U}\rightarrow\mathbb{W}$ of symbols, where concatenation
on domains $\mathbb{U}^*$ and $\mathbb{W}^*$ is defined, denote
its extension onto strings as $f^*:\mathbb{U}^*\ni x_1x_2...x_m\mapsto
f(x_1)f(x_2)...f(x_m)\in\mathbb{W}^*$. For grammars
$G_i=(\alpha_{i1},\alpha_{i2},...,\alpha_{in_i})$, $i=1,2$, define
joining
\begin{align*}
G_1\oplus G_2:=(A_2A_{n_1+2},\,
&H_1^*(\alpha_{11}),H_1^*(\alpha_{12}),...,H_1^*(\alpha_{1n_1}),
\\
&H_2^*(\alpha_{21}),H_2^*(\alpha_{22}),...,H_2^*(\alpha_{2n_2})),
\end{align*}
where $H_1(A_j):=A_{j+1}$ and $H_2(A_j):=A_{j+n_1+1}$ for nonterminals
and $H_1(x):=H_2(x):=x$ for terminals $x\in\mathbb{X}$.  

\begin{definition}
$B:\mathcal{G}\rightarrow\mathbb{X}^+$ is
a~\emph{local grammar encoder} if
\begin{align}
\label{LocalCoder}
B(G)=B_\text{S}^*(B_\text{N}(G)), 
\end{align}
where:
\begin{LaTeXenumerate}
\item function
$B_\text{N}:\mathcal{G}\rightarrow(\klam{0}\cup\mathbb{N})^*$ encodes
grammars as strings of natural numbers so that the encoding of grammar
$G=(\alpha_1,\alpha_2,...,\alpha_n)$ is string
$$B_\text{N}(G):=F_1^*(\alpha_1)DF_2^*(\alpha_2)D...DF_n^*(\alpha_n)(D+1),$$
which employs relative indexing $F_i(A_j):=D+1+j-i$ for nonterminals and
identity transformation $F_i(x):=x$ for terminals
$x\in\mathbb{X}=\klam{0,1,...,D-1}$,
\item $B_\text{S}$  is any function of form $B_\text{S}:\klam{0}\cup\mathbb{N}
 \rightarrow\mathbb{X}^+$ (for technical purposes, not necessarily
 an injection)---we will call $B_\text{S}$ the natural number encoder.
\end{LaTeXenumerate}
\end{definition}
Indeed, local encoders are adapted to joining operation $\oplus$.
For instance, if $B(G_i)=u_iB_\text{S}(D+1)$ for some grammars $G_i$,
$i=1,2$, then $B(G_1\oplus G_2)=
B_\text{S}(D+2)B_\text{S}(D+2+\voc{G_1})
B_\text{S}(D)u_1B_\text{S}(D)u_2B_\text{S}(D+1)$.

There exist many prefix-free local encoders.  Obviously, set
$B_\text{N}(\mathcal{G})$ itself is prefix-free.  Therefore, encoder
(\ref{LocalCoder}) is prefix-free (and uniquely decodable) if
$B_\text{S}$ is also prefix-free, i.e., if $B_\text{S}$ is an
injection and set $B_\text{S}(\klam{0}\cup\mathbb{N})$ is prefix-free.

\subsection{Encoder-induced grammar lengths}
\label{ssecLengths}

Let us generalize the concept of grammar length.
\begin{definition}
For a~grammar encoder $B$, function $|B(\cdot)|$
will be called the \emph{$B$-induced grammar length}.
\end{definition}
For example, Yang-Kieffer length  $\abs{\,\cdot\,}$ is
$B$-induced for a~local grammar encoder
$B=B_\text{S}^*(B_\text{N}(\cdot))$, where
\begin{align}
\label{YKCoder}
\text{$B_\text{S}(x)=\lambda$ for $x\in\klam{D,D+1}$ 
and $B_\text{S}(x)\in\mathbb{X}$ else}.
\end{align}

In the same spirit, we can extend the idea of the smallest grammar
with respect to the Yang-Kieffer length, discussed in
\cite{CharikarOthers05}.  Subclass $\mathcal{J}\subset\mathcal{G}$ of
admissible grammars will be called \emph{sufficient} if there exists
a~grammar transform $\Gamma:\mathbb{X}^+\rightarrow\mathcal{J}$, i.e.,
if $\mathcal{J}\cap \mathcal{G}(w)\not=\emptyset$ for all
$w\in\mathbb{X}^+$. Conversely, we will call grammar transform
$\Gamma$ a~$\mathcal{J}$-grammar transform if
$\Gamma(\mathbb{X}^+)\subset\mathcal{J}$.
\begin{definition}
  For grammar length $\aabs{\cdot}$, $\mathcal{J}$-grammar transform
  $\Gamma$ will be called \emph{$(\aabs{\cdot},\mathcal{J})$-minimal
    grammar transform} if $\aabs{\Gamma(w)}\le \aabs{G}$ for all
  $G\in\mathcal{G}(w)\cap \mathcal{J}$ and $w\in\mathbb{X}^+$.
\end{definition}
\begin{definition}
Code $B(\Gamma(\cdot))$ will be called
\emph{$(B,\mathcal{J})$-minimal} if $\Gamma$ is
$(\aabs{\cdot},\mathcal{J})$-minimal for a~$B$-induced grammar length
$\aabs{\cdot}$.
\end{definition}
\begin{definition}
For a~grammar length $\aabs{\cdot}$, grammar subclasses
$\mathcal{J},\mathcal{K}\subset\mathcal{G}$ are called
\emph{$\aabs{\cdot}$-equivalent} if
$$\min_{G\in\mathcal{G}(w)\cap\mathcal{J}}
\aabs{G}=\min_{G\in\mathcal{G}(w)\cap\mathcal{K}} \aabs{G}\qquad \text{for
  all $w\in\mathbb{X}^+$}.$$
\end{definition}

\subsection{Subclasses of grammars}
\label{ssecGrammars}

In section \ref{secVocabulary}, we will bound the excess lengths for
$(B,\mathcal{J})$-minimal codes, where $B$ are local encoders and
$\mathcal{J}$ are some sufficient subclasses. In subsection
\ref{ssecUniversal}, we will show that several of these codes are
universal. Prior to this, we have to define some necessary subclasses
of grammars.

First, we will say that $(\alpha_1,\alpha_2,...,\alpha_n)$ is
a~\emph{flat grammar} if $\alpha_i\in \mathbb{X}^+$ for $i>1$. The set
of flat grammars will be denoted as $\mathcal{F}$. Symbol
$\mathcal{D}_k\subset\mathcal{F}$ will denote the class of
\emph{$k$-block interleaved grammars}, i.e., flat grammars
$(\alpha_1,\alpha_2,...,\alpha_n)$, where $\alpha_i\in \mathbb{X}^k$
for $i>1$. On the other hand, $\mathcal{B}_k\subset\mathcal{D}_k$ will
stand for the set of \emph{$k$-block grammars}, i.e., $k$-block
interleaved grammars $(uw,\alpha_2,...,\alpha_n)$, where string
$u\in(\klam{A_{2},A_{3},...,A_n})^*$ contains occurrences of all
$A_{2},A_{3},...,A_n$ and string $w\in\mathbb{X}^*$ has length
$|w|<k$, cf.\ \cite{NeuhoffShields98}. Of course, classes
$\mathcal{B}_k$, $\mathcal{D}_k$, $\mathcal{B}:=\bigcup_{k\ge 1}
\mathcal{B}_k$, $\mathcal{D}:=\bigcup_{k\ge 1} \mathcal{D}_k$, and
$\mathcal{F}$ are sufficient.

Next, grammar $(\alpha_1,\alpha_2,...,\alpha_n)$ is called
\emph{irreducible} if
\begin{LaTeXenumerate}
\item each string $\alpha_i$ has a~different expansion
  $\dzi{\alpha_i}_G$ and satisfies $\abs{\alpha_i}>1$,
\item each secondary nonterminal appears in string
  $\alpha_1\alpha_2...\alpha_n$ at least twice,
\item each pair of consecutive symbols in strings
  $\alpha_1,\alpha_2,...,\alpha_n$ appears at most once at
  nonoverlapping positions \cite{KiefferYang00}.
\end{LaTeXenumerate}
The set of irreducible grammars will be denoted as $\mathcal{I}$. Any
$\mathcal{I}$-grammar transform is asymptotically compact
\cite{KiefferYang00} so it yields a~universal code when combined with
grammar encoder $B_\text{YK}$.

Starting with any grammar $G_1\in\mathcal{G}(w)$, one can construct an
irreducible grammar $G_2\in\mathcal{G}(w)$ by applying a~sequence of
certain reduction rules until the local minimum of functional
$2\abs{\,\cdot\,}-\voc{\cdot}$ is achieved \cite{KiefferYang00}. This
leads to the following lemma.
\begin{lemma}
\label{theoLengthIrred}
Classes $\mathcal{I}$ and $\mathcal{G}$ are
$\abs{\,\cdot\,}$-equivalent.
\end{lemma}
\begin{proof}
  The only reduction rule applicable to a~grammar minimizing
  $\abs{\,\cdot\,}$ is the introduction of a~new nonterminal denoting
  a~pair of symbols which appears exactly twice on the right-hand side
  of the grammar, cf. section VI in \cite{KiefferYang00}. This
  reduction conserves Yang-Kieffer length.
\end{proof}

Additionally, we will say that grammar
$(\alpha_1,\alpha_2,...,\alpha_n)$ is \emph{partially irreducible} if
it satisfies conditions (i) and (ii) of irreducibility, as well as,
each pair of consecutive symbols in string $\alpha_1$ appears at most
once at nonoverlapping positions.
Let $\mathcal{P}$ stand for the set of partially irreducible
grammars. Of course, $\mathcal{I}\subset\mathcal{P}\subset\mathcal{G}$
and $\mathcal{P}$ is sufficient.  

Although $\mathcal{F}\cap\mathcal{P}$ and $\mathcal{F}$ are not
$\abs{\,\cdot\,}$-equivalent, class $\mathcal{F}\cap\mathcal{P}$ is
sufficient and relates to $\mathcal{F}$ partially like $\mathcal{I}$
relates to $\mathcal{G}$. Some $\mathcal{F}\cap\mathcal{P}$-grammar
transform $\Gamma$ is a~modification of the longest matching
$\mathcal{I}$-grammar transform \cite{KiefferYang00,CharikarOthers05}.
In order to compute $\Gamma(w)$, we start with grammar
$\klam{A_1\rightarrow w}$ and we replace iteratively the longest
repeated substrings $u$ in the start symbol definition with new
nonterminals $A_i\rightarrow u$ until there is no repeat of length
$|u|\ge 2$. $\Gamma(w)$  is  the modified grammar.

\subsection{Universal codes for local encoders}
\label{ssecUniversal}

Neuhoff and Shields proved that any
$(B_\text{NS},\mathcal{B})$-minimal code is universal for some encoder
$B_\text{NS}$ and the class of block grammars $\mathcal{B}$
\cite{NeuhoffShields98}.  Encoder $B_\text{NS}$ resembles a~local
encoder.  The main difference is encoding nonterminals $A_i$ as
strings of length $\floor{\log_D \voc{G}}+ 1$ rather than strings of
length $|B_\text{S}(D+i)|$. Therefore we can establish the following
proposition.
\begin{theorem}
\label{theoUni}
Let $B_\text{S}$ be such a~prefix-free natural number encoder
that $|B_\text{S}(\cdot)|$ is growing and
\begin{align}
\label{UniCoder}
\limsup_{n\rightarrow\infty} |B_\text{S}(n)|/\log_D n =1.
\end{align}
Then for any sufficient subclass of grammars
$\mathcal{J}\supset\mathcal{B}$, every
$(B_\text{S}^*(B_\text{N}(\cdot)),\mathcal{J})$-minimal code $C$ is
universal, that is, $\lim_{n} H^C(n)/n=h$ and $\limsup_{n}
K^C(X_{1:n})/n\le h$ almost surely for every stationary process
$(X_k)_{k\in\mathbb{Z}}$.
\end{theorem}
\begin{proof}
Consider $\mathcal{B}_k$-grammar transforms $\Gamma_k$. 
For $\epsilon>0$ and stationary process
$(X_k)_{k\in\mathbb{Z}}$ with entropy rate $h$, let $k(n)$ be the largest
integer $k$ satisfying $k2^{k(H+\epsilon)}\le n$. We have
\begin{align*}
\limsup_{n\rightarrow\infty} \max_{w\in\mathbb{X}^n} 
\frac{\log_D \voc{\Gamma_{k(n)}(w)}}{k(n)} &\le h+2\epsilon
,
\\
\lim_{n\rightarrow\infty} 
\sred{\voc{\Gamma_{k(n)}(X_{1:n})}}\cdot k(n)/n  &= 0
,
\\
\lim_{n\rightarrow\infty} 
\voc{\Gamma_{k(n)}(X_{1:n})}\cdot  k(n)/n &= 0
\text{ almost surely, cf.\ \cite{NeuhoffShields98}}
.
\end{align*}
Since $\lim_n k(n)=\infty$, a~$(B,\mathcal{J})$-minimal
code is universal if
\begin{align*}
|B(\Gamma_k(w))|\le \alpha k\voc{\Gamma_{k}(w)} + \gamma(k)\frac{n}{k}
\log_D \voc{\Gamma_{k}(w)},
\end{align*}
where $\alpha >0$ and $\lim_k \gamma(k)=1$. In particular, this
inequality holds for (\ref{LocalCoder}), (\ref{UniCoder}), and growing
$|B_\text{S}(\cdot)|$.
\end{proof}

The prefix-free natural number encoder $B_\text{S}$ satisfying
(\ref{UniCoder}) can be chosen, e.g., as the $D$-ary representation
$\omega:\mathbb{N}\rightarrow\mathbb{X}^*$ \cite{Elias75},
$|\omega(n)|=\ell(n)$, where $$\ell(n) :=
\begin{cases}     
  1                   &\text{if } n < D, \\
  \ell(\floor{\log_D n}) + \floor{\log_D n} + 1  &\text{if } n \ge D.  
\end{cases}$$
Alternatively, we can use the $D$-ary representation
$\delta:\mathbb{N}\rightarrow\mathbb{X}^*$ \cite{Elias75},
$|\delta(n)|= 
1
+2\floor{\log_D (1+\floor{\log_D n})} 
+\floor{\log_D n}$.

\section{Bounds involving the vocabulary size}
\label{secVocabulary}

We will derive several inequalities for the vocabulary size of certain
minimal grammar-based codes.  Frankly speaking, code universality is
irrelevant for the proofs.  It is important, however, that the codes
use the local grammar encoders.

\subsection{Upper bounds for the  excess lengths}
\label{ssecUpperExcess}

We will begin with defining several operations on grammars.  For
strings $u,v\in\mathbb{X}^*$ with $n=\abs{u}$, $m=\abs{v}$, and $w=uv$, 
define the \emph{left} and \emph{right croppings} of grammar
$G=(\alpha_1,\alpha_2,...,\alpha_n)\in\mathcal{G}(w)$ as
\begin{align*}
\mathbb{L}_n G:=(x_Ly_L,\alpha_2,...,\alpha_n)\in\mathcal{G}(u),
\\
\mathbb{R}_m G:=(y_Rx_R,\alpha_2,...,\alpha_n)\in\mathcal{G}(v),
\end{align*}
where exactly one of the following conditions holds:
\begin{LaTeXenumerate}
\item $\alpha_1=x_Lx_R$ and $y_Ly_R=\lambda$, 
\item $\alpha_1=x_LA_ix_R$ for some nonterminal $A_i$, $2\le i\le n$,
with expansion $\dzi{A_i}_G=y_Ly_R$.
\end{LaTeXenumerate}

Next, for  $G=(\alpha_1,\alpha_2,...,\alpha_n)$, define its
\emph{flattening} $\mathbb{F}G:=
(\alpha_1,\dzi{\alpha_2}_G,\dzi{\alpha_3}_G,...,\dzi{\alpha_n}_G)$.
The secondary part of the grammar will be denoted as $\mathbb{S}G:=
(\lambda,\alpha_2,\alpha_3,...,\alpha_n)$.  Additionally, we will use
a~notation for the maximal length of a~nonoverlapping repeat in string
$w\in\mathbb{X}^*$, i.e.,
\begin{align*}
\mathbf{L}(w):=\max_{u,x,y,z\in\mathbb{X}^*:\, w=xuyuz} |u|.
\end{align*}

Now we can generalize Theorem 3 from \cite{Debowski06}. We will show
that the lengths of some minimal codes are almost subadditive.
Moreover, the excess lengths are dominated by the vocabulary size
multiplied by the length of the longest repeat.
\begin{theorem}
\label{theoUpper}
Let $B$ be local encoder (\ref{LocalCoder}). Introduce constants
\begin{align*}
W_m&:=\max_{0\le n\le D+2+m}|B_\text{S}(n)|.
\end{align*}
Let $\Gamma$ be a~$(\aabs{\cdot},\mathcal{J})$-minimal grammar
transform for the $B$-induced grammar length $\aabs{\cdot}$.  Consider
code $C=B(\Gamma(\cdot))$, strings $u,v,w\in\mathbb{X}^+$, and
a~grammar class $\mathcal{K}$ which is $\aabs{\cdot}$-equivalent to
$\mathcal{J}$.
\begin{LaTeXenumerate}
\item If $G_1,G_2\in\mathcal{J}\implies G_1\oplus G_2\in\mathcal{K}$
then
\begin{align}
\label{UpperLower}
\abs{C(u)}+\abs{C(v)}-\abs{C(uv)} \ge -3W_0 - W_{\voc{\Gamma(u)}}.
\end{align}
\item If $G\in\mathcal{J}\implies \mathbb{L}_n G,\, \mathbb{R}_n
  G\in\mathcal{K}$ for all valid $n$ then
\begin{align}
\label{UpperLeftRight}
\hspace{-2em}
\abs{C(u)},\, \abs{C(v)}
&\le \abs{C(uv)} + W_0\mathbf{L}(uv),
\\
\label{UpperUpper}
\hspace{-2em}
\abs{C(u)}+\abs{C(v)}-\abs{C(uv)} 
&\le \aabs{\mathbb{S}\Gamma(uv)}+W_0\mathbf{L}(uv).
\end{align}
\item If $G\in\mathcal{J}\implies \mathbb{F}G\in\mathcal{K}$ then
\begin{align}
\label{UpperVoc}
\aabs{\mathbb{S}\Gamma(w)}+W_0\mathbf{L}(w)\le  W_0\voc{\Gamma(w)}(1+\mathbf{L}(w)).
\end{align}
\end{LaTeXenumerate}
\emph{Remark 1:} In particular, (\ref{UpperLower}) holds for
$\mathcal{J}=\mathcal{G},\mathcal{P},\mathcal{I}$ while inequalities
(\ref{UpperLeftRight})--(\ref{UpperVoc}) hold for
$\mathcal{J}=\mathcal{G},\mathcal{P},\mathcal{I},\mathcal{F},\mathcal{D},\mathcal{D}_k$.
Moreover, (\ref{UpperUpper}) and (\ref{UpperVoc}) imply together bound
\begin{align}
\label{UpperVocII}
\hspace{-0.5em}
\abs{C(u)}+\abs{C(v)}-\abs{C(uv)}
&\le  W_0\voc{\Gamma(uv)}(1+\mathbf{L}(uv)),
\end{align}
which we have mentioned in the introduction.
\\
\emph{Remark 2:} Theorem 3 in \cite{Debowski06} is a~restriction of
Theorem \ref{theoUpper} to $B_\text{S}$ given by (\ref{YKCoder}) and
$\aabs{\cdot}$ equal to Yang-Kieffer length $\abs{\,\cdot\,}$.
\end{theorem}

 \begin{proof}
\begin{LaTeXenumerate}
\item The result is implied by $\aabs{\Gamma(uv)}\le
  \aabs{\Gamma(u)\oplus\Gamma(v)}$ and $$\aabs{G_1\oplus G_2}
  \le\aabs{G_1}+\aabs{G_2}+|B_\text{S}(D+2+\voc{G_1})|+3W_0,$$
where $G_1=\Gamma(u)$ and $G_2=\Gamma(v)$.
\item Set $n=\abs{u}$, $m=\abs{v}$, and $w=uv$. The inequalities follow from
\begin{align*}
\aabs{\Gamma(w)}+W_0\mathbf{L}(w)&\ge
  \aabs{\mathbb{L}_n\Gamma(w)}\ge \aabs{\Gamma(u)}, 
\\
\aabs{\Gamma(w)}+W_0\mathbf{L}(w)&\ge
  \aabs{\mathbb{R}_m\Gamma(w)}\ge \aabs{\Gamma(v)}, 
\end{align*}
and
  $$\aabs{\mathbb{L}_n\Gamma(w)}+\aabs{\mathbb{R}_m\Gamma(w)}\le
  \aabs{\Gamma(w)}+ \aabs{\mathbb{S}\Gamma(w)}+W_0\mathbf{L}(w).$$
\item The thesis is entailed by $\aabs{\mathbb{S}\Gamma(w)}\le
  \aabs{\mathbb{S}\mathbb{F}\Gamma(w)}$ and
  $\aabs{\mathbb{S}\mathbb{F}\Gamma(w)}\le
  W_0\okra{\voc{\Gamma(w)}-1}(1+\mathbf{L}(w))+W_0$.
\end{LaTeXenumerate}
\end{proof}

\subsection{Lower bounds for the  excess lengths}
\label{ssecLowerExcess}

For Yang-Kieffer length function, the excess lengths can be lower-bounded
by another quantity related to vocabulary size.  Firstly, for grammars
$G_i=(\alpha_{i1},\alpha_{i2},...,\alpha_{in_i})$, $i=1,2$, denote the
number of their common nonterminal expansions 
\begin{align*}
\voc{G_1;G_2}:= \card \bigcap_{i=1,2}
\klam{\dzi{\alpha_{i2}}_{G_i},\dzi{\alpha_{i3}}_{G_i},...,
\dzi{\alpha_{in_i}}_{G_i}}
\end{align*}
and introduce a~new kind of grammar joining
\begin{align*}
G_1\otimes G_2:=(\alpha_{11}\alpha_{21},\,
&Q_1^*(\alpha_{12}),...,Q_1^*(\alpha_{1n_1}),
\\
&Q_2^*(\alpha_{22}),...,Q_2^*(\alpha_{2n_2})),
\end{align*}
where $Q_1(A_j):=A_{j}$ and $Q_2(A_j):=A_{j+n_1-1}$ for nonterminals
and $Q_1(x):=Q_2(x):=x$ for terminals $x\in\mathbb{X}$.  

Recall also Grammar Reduction Rule 5 from \cite{KiefferYang00}, which
deletes useless nonterminals from the grammar and, for all
nonterminals sharing the same expansion, substitutes one of them. Let
$\mathbb{I}G$ be the result of applying the rule to grammar $G$.
\begin{theorem}
\label{theoLower}
Let $\Gamma$ be a~$(\abs{\,\cdot\,},\mathcal{J})$-minimal grammar
transform.  If
$G_1,G_2\in\mathcal{K}\implies \mathbb{I}G_1, G_1\otimes
G_2\in\mathcal{K}$ for some grammar class $\mathcal{K}$ being
$\abs{\,\cdot\,}$-equivalent to $\mathcal{J}$ then
\begin{align}
\label{LowerLower}
\abs{\Gamma(u)}+\abs{\Gamma(v)}-\abs{\Gamma(uv)}
&\ge 
\voc{\Gamma(u);\Gamma(v)}.
\end{align}
\emph{Remark:} In particular, (\ref{LowerLower}) holds for
$\mathcal{J}=\mathcal{G},\mathcal{P},\mathcal{I},\mathcal{F},\mathcal{D}_k$.
\end{theorem}
\begin{proof}
  Since $\mathcal{K}$ is closed against operation $\mathbb{I}$, there
  exist $G_1\in\mathcal{K}\cap\mathcal{G}(u)$ and
  $G_2\in\mathcal{K}\cap\mathcal{G}(v)$ such that
  $\abs{G_1}=\abs{\Gamma(u)}$, $\abs{G_2}=\abs{\Gamma(v)}$, and
  $\mathbb{I}G_i=G_i$. Hence $\abs{\alpha_{ij}}\ge 1$ for
  $(\alpha_{i1},\alpha_{i2},...,\alpha_{in_i})=G_i$ and, consequently,
\begin{align}
\abs{\mathbb{I}(G_1\otimes G_2)} 
&\le 
\abs{G_1\otimes G_2} -
\voc{G_1;G_2}\min_{ij} \abs{\alpha_{ij}}
\nonumber
\\
\label{PreLowerLower}
&\le
\abs{G_1\otimes G_2} -
\voc{G_1;G_2}
.
\end{align}
Notice that $\abs{G_1\otimes G_2}=\abs{G_1}+\abs{G_2}$.  Thus
(\ref{LowerLower}) follows from (\ref{PreLowerLower}) and from
$\abs{\Gamma(uv)}\le \abs{\mathbb{I}(G_1\otimes G_2)}$.
\end{proof}

The next proposition suggests that the size of common vocabulary
$\voc{\Gamma(u);\Gamma(v)}$ for irreducible grammar transforms may grow
quite fast with the length of strings $u$ and $v$.
\begin{theorem}
\label{theoIrred}
\begin{LaTeXenumerate}
\item 
If $\Gamma$ is a~$\mathcal{F}\cap\mathcal{P}$-grammar transform then
\begin{align}
  \label{PIrred}
  \voc{\Gamma(w)}\mathbf{L}(w)> \sqrt{\abs{\Gamma(w)}/2}-D-1.
\end{align}
\item
If $\Gamma$ is an $\mathcal{I}$-grammar transform then
\begin{align}
  \label{Irred}
  \voc{\Gamma(w)}> \sqrt{\abs{\Gamma(w)}/2}-D-1.
\end{align}
\end{LaTeXenumerate}
\emph{Remark:} Bound (ii) was mentioned in \cite{Debowski06b}.
\end{theorem}
\begin{proof}
  Write $G=\Gamma(w)$ and $V=\voc{\Gamma(w)}$ for brevity.
  Notice that $x+a+1> \sqrt{y/2}$ follows from 
  $(y-x)/2 \le (x+a)^2$ for $x,y,a\ge 0$.
  \begin{LaTeXenumerate}
  \item At the every second position of the start symbol definition of $G$,
    a~pair of symbols can occur only once.  Thus (\ref{PIrred})
    follows by $[\abs{G}-V\mathbf{L}(w)]/2 \le \okra{V+D}^2 \le
    \okra{V\mathbf{L}(w)+D}^2$.
  \item In this case, any pair of symbols occurs at most once at the
    every second position of all right-hand sides of $G$. Hence,
    $(\abs{G}-V)/2 \le \okra{V+D}^2$, which implies (\ref{Irred}).
  \end{LaTeXenumerate}
\end{proof}

\section{Conclusion}
\label{secConclude}

We have shown that the vocabulary size of certain minimal universal
grammar-based codes is greater than the excess code length divided by
the length of the longest repeated substring $\mathbf{L}(\cdot)$.
Recall that $\mathbf{L}(X_{1:n})$ cannot be upper-bounded almost
surely by a~universal function $o(n)$ for a~block of $n$ symbols drawn
from an arbitrary stationary stochastic process \cite{Shields92b}.
Nevertheless, $\mathbf{L}(X_{1:n})=O(\log n)$  if
$(X_i)_{i\in\mathbb{Z}}$ is a~finite-energy process \cite{Shields97}.
Hence, an extended Hilberg hypothesis \cite{Hilberg90}, stating that
a~good model for texts in natural languages is a~finite-energy process
with excess entropy $E(n)\asymp\sqrt{n}$, seems consistent with
observations asserting that vocabulary size for certain text
compressions is $\Omega(\sqrt{n}/\log n)$ where $n$ is the text length
\cite[Figure 3.12 (b), p.\ 69]{NevillManning96}.

While some premises appealing to ergodic decomposition make Hilberg's
hypothesis plausible even without the evidence of grammar-based
compression \cite{Debowski06c}, there remains an important theoretical
problem.  Can we use the vocabulary size or the excess length of
a~grammar-based code to estimate excess entropy accurately?
Inequality (\ref{DiffECE}) gives a~lower bound for 
$E^C(n)-E(n)$ but the upper bounds are less recognized.  Although
$\abs{E^C(n)-E(n)}=O(\log n)$ when the length of code $C$ equals
prefix algorithmic complexity and block distribution $P(X_{1:n})$ is
recursively computable \cite{Debowski06c,GrunwaldVitanyi03}, some
results in ergodic theory indicate that there is no universal bound
for $\abs{E^C(n)-E(n)}$ in the class of stationary processes
\cite{Debowski06c,Shields93}.

Simpler arguments could be used to infer that difference $E^C(n)-E(n)$
is large for certain codes and stochastic processes. Consider
compressing a~memoryless source with entropy rate $h>0$. We have
$E(n)=0$. On the other hand, let code $C$ be formed by a~local encoder
satisfying (\ref{UniCoder}) and an irreducible transform $\Gamma$.
Then $E^C(n)= \Omega(\sqrt{hn/\log n})$ would be implied by
Theorems \ref{theoLower} and \ref{theoIrred} if relation
$\voc{\Gamma(X_{1:n});\Gamma(X_{n+1:2n})}\asymp \voc{\Gamma(X_{1:n})}$
held.

Let us notice that the bound for $E^C(n)$ conjectured for memoryless
sources and irreducible grammar-based codes is almost the same as the
inequality established for general minimal codes and sources with
$E(n)\asymp\sqrt{n}$.  This should not obscure the fact that there is
a~huge variation of vocabulary size for different information sources
and a~fixed code \cite{Debowski06b}, an empirical fact not yet fully
understood theoretically.

\section*{Acknowledgment}
This work was supported by the Australian Research Council, grant no.\
DP0210999, during the author's visit to the University of New South
Wales, Sydney, Australia.  The author wishes to thank to Prof. Arthur
Ramer of the UNSW.

\newpage

\end{document}